# A study of random resistor-capacitor-diode networks to assess the electromagnetic properties of carbon nanotube filled polymers

D. S. Bychanok,[1a] A. G. Paddubskaya,[1] P. P. Kuzhir,[1] S. A. Maksimenko,[1] C. Brosseau,[2] J. Macutkevic,[3] and S. Bellucci[4]

[1]*Research Institute for Nuclear Problems BSU, 11 Bobruiskaya str., 220030, Minsk, Belarus*
[2]*Université de Brest, Lab-STICC, CS 93837, 6 avenue Le Gorgeu, 29238 Brest Cedex 3, France*
[3]*Vilnius University, Sauletekio al. 9, LT-00122 Vilnius, Lithuania*
[4]*INFN-Laboratori Nazionali di Frascati, 00044 Frascati, Italy*

We determined the frequency dependent effective permittivity of a large ternary network of randomly positioned resistors, capacitors, and diodes. A linear circuit analysis of such systems is shown to match the experimental dielectric response of single-walled carbon nanotube (SWCNT) filled polymers. This modeling method is able to reproduce the two most important features of SWCNT filled composites, i.e. the low frequency dispersion and dipolar relaxation. As a result of the modeling important physical conclusion proved by the experimental data was done: the low frequency behavior of SWCNT-filled polymer composites is mostly caused by the fraction of semiconducting SWCNTs.

Filling polymers with carbonaceous nanoparticles is known to have profound effects on their electromagnetic wave transport and mechanical properties with examples ranging from carbon black to carbon nanotubes (CNTs).[1-8] A great deal of current research focuses on plastic electronics.[9] In some respects, soft materials filled by CNTs are ideally suited for electronic applications because of potentially interesting emergent interfacial phenomena between graphene and polymer chains. Experimental progress in CNT based polymer nanocomposites preparation has steadily advanced recently due largely to the control of the elasticity network during their fabrication.[1,10-12] Since the processing may modify the filler mesostructure, the dielectric properties of heterostructures are not yet well understood, not least owing to a lack of well-controlled systems.

A theme common to disordered materials is emergent behaviour under ac excitation. A typical example is the low frequency dispersion (LFD) which is widely observed in composite materials. While research on the electrical and electromagnetic responses of networks of randomly positioned resistors (R) and capacitors (C) has a long history in materials science, it continues to be important for understanding the properties of strongly correlated materials in which the charge and wave behaviors are strongly related to the heterogeneous microstructure.[13-16] Depending on CNT diameter and chirality, the electrical conductivity of CNTs can show metallic or semiconducting behavior. Typically, one third of single-walled (SW) CNT nanocylinders have metallic behavior and the rest are semiconducting.[17] Consequently, we choose to introduce diodes (D) for modeling the microstructure that is a ternary network of interconnected conductive, dielectric, and semiconducting phases. It is worth noting that the incorporation of inductance elements can be also considered. Preliminary simulations performed with reported inductances values for CNTs and CNT bundles[18] did not affect significantly the low frequency electromagnetic response of CNT-based composites. Therefore, we restrict our attention to RCD networks.

In this Letter, we present a combined RCD theory and experimental comparison study of the electromagnetic properties of SWCNT-filled polymers in the frequency range from $10^2$ to $10^6$ Hz. We show that such a model is able to describe the LFD of supercolative composites and the dipolar relaxation[19] in subpercolative composites. We show that our results are consistent with the universal dielectric response that is an emergent property arising out of the random nature of the composite.[20] In order to demonstrate that our approach is well-founded, we reproduce some of the known results of earlier experimental studies[10,21,22] devoted to a broadband analysis of the complex effective permittivity spectra of SWCNT-filled polymers.

Keeping in mind the picture sketched in Fig. 1, let us consider a two-dimensional RCD network, which for simplicity we assume to be finite and square. The network considered is a square 41 component lattice. Identical responses have been also obtained for much larger networks (not presented). There are related studies on microstructural networks of R and C regions modeling the electrical response of heterostructures.[13,14] Kirchhoff's circuit laws allow us to express the electrical response of such a structure by a system of linear equations where the unknowns are the currents in the network. Solving of this system yields the frequency dependence of the impedance.

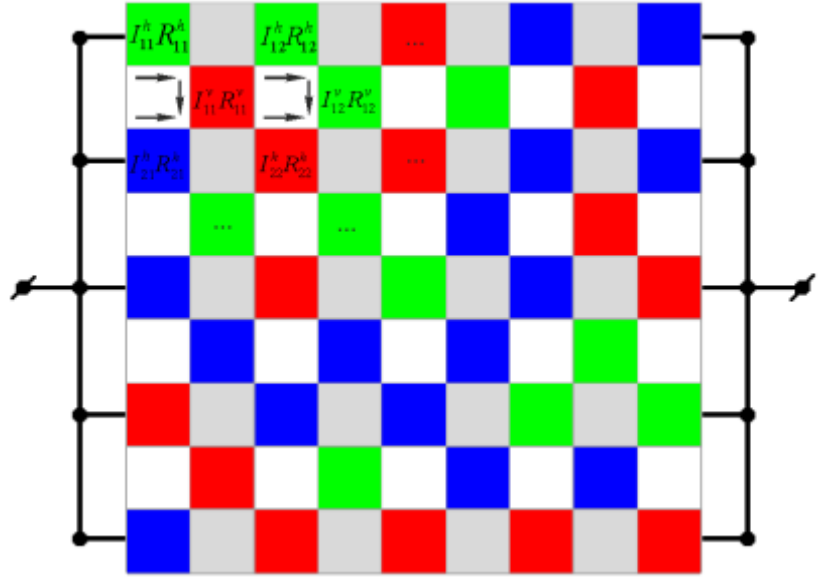

FIG 1: (color online) Illustration of an example ternary network (two dimensional square checkerboard) comprising a disordered mixture of resistors (shown in red), capacitors (shown in blue), and diodes (shown in green). Electrical contacts are shown in grey, while porosity is shown in white.

We consider the square checkerboard structure composed of $n$ horizontal rows. Each row has $m$ elements and is connected to its neighbor by $m-1$ elements. The system has $N = nm + (n-1)(m-1)$ elements, $N$ is also the number of unknown currents that are needed to determine the frequency dependent complex impedance of the system given by $\hat{Z} = 1/\hat{I}$. Here, $\hat{I}$ is the total complex current through the system by applying a voltage drop asserted to electrodes (set to 1 V). The impedance of each intrinsic element in the system is $R$ for each resistor, $1/i\omega C_0$ for each capacitor, and $1/(i\omega)^k C_1$ for each diode.[23] Here $i = \sqrt{-1}$, $\omega$ is the angular frequency of the electric field, while $C_0$ and $C_1$ are scaling factors. The unknown currents can be quantified as follows. We first assume that all currents in each row are oriented rightward and downward for each column (see the arrows in Fig.1). We define two matrices $\hat{R}^h$ and $\hat{R}^v$, respectively of dimensions $n \times m$ and $(n-1) \times (m-1)$ with elements $R_{ij}^h$ and $R_{ij}^v$, corresponding to the elements of the RCD network through

$$R_{ij}^h, R_{ij}^v = \delta_{-1\gamma} R + \delta_{0\gamma} \frac{1}{i\omega C_0} + \delta_{1\gamma} \frac{1}{(i\omega)^k C_1}, \quad (1)$$

where $\delta_{i\gamma}$ is the Kronecker delta and $\gamma$ is a random variable that takes a value set to -1, 0, or 1, corresponding respectively to resistors, capacitors and diodes. Thus, the relative content of diodes, resistors and capacitors in the system can be tuned by the $\gamma$ distribution. In this work we investigate random networks by considering a uniform $\gamma$ distribution leading to rough equal amounts of R, C, and D.

Now we can define two unknown current matrices $\hat{I}^h$ and $\hat{I}^v$ with dimensions $n \times m$ and $(n-1) \times (m-1)$. Application of the Kirchhoff's circuit laws for each row, each square part and each node of the system leads to the following set of coupled equations

$$\begin{cases} \sum_{j=1}^{m} R_{ij}^h I_{ij}^h = 1 \\ R_{ij}^h I_{ij}^h + (1-\delta_{jm})R_{ij}^v I_{ij}^v - R_{i+1j}^h I_{i+1j}^h - (1-\delta_{j1})R_{ij-1}^v I_{ij-1}^v = 0 \\ I_{ij}^h - I_{ij+1}^h + (1-\delta_{i1})I_{i-1j}^v - (1-\delta_{in})I_{ij}^v = 0. \end{cases} \quad (2)$$

An important aspect of Eq. (2) is that it represents an overdetermined system since the $m \times (2n-1)$ equations contain $N$ unknown variables. Solving for Eq. (2) relative to the unknown $\hat{I}^h$ and $\hat{I}^v$ allows calculation of the total current $I = \sum_{i=1}^{n} I_{i,1}^h = \sum_{i=1}^{n} I_{i,m}^h$ through the system and of the impedance. As tests, we performed these calculations for the cases of random RC and RCD networks with no percolation path for diodes and found that the results are consistent with earlier studies.[13,14] A program in Python and the solvers module in SymPy were used for solving Eq.(2).

In our quasistatic simulations, we have assumed that the composite structure has a surface $S = 2$ cm$^2$ between the two electrodes separated by a distance $d$=1 mm. The external electromagnetic radiation follows a $\exp(i\omega t)$ time dependence. The quasistatic approximation means that all length scales that characterize the inhomogeneities in the material medium and the skin depth in case of a conducting inclusion must be much smaller than the wavelength of radiation. If we set $C = \varepsilon \varepsilon_0 S/d = \varepsilon C_c$, the effective permittivity and conductivity spectra are given by

$$\varepsilon(\omega) = \varepsilon' + i\varepsilon'' = \frac{1}{i\omega C_c Z(\omega)}, \sigma(\omega) = \omega \varepsilon_0 \varepsilon''. \quad (3)$$

Note that Eq.(3) can be considered in the numerical studies when the capacitor percolative path is present in the network. This is consistent with real experimental studies of polymer composites with low CNT concentration, typically <2 wt.%. The specific question we are addressing in this Letter is the following: given a set of three phases R, C, and D and a physical network model, how does the complex impedance and corresponded complex effective permittivity vary with frequency? We start with an example of calculations which are shown in Fig.2, where $R = 10^3$ k$\Omega$, $C_0 = 0.1$ nF, $C_1$=50 nF, and $k = 0.1$.

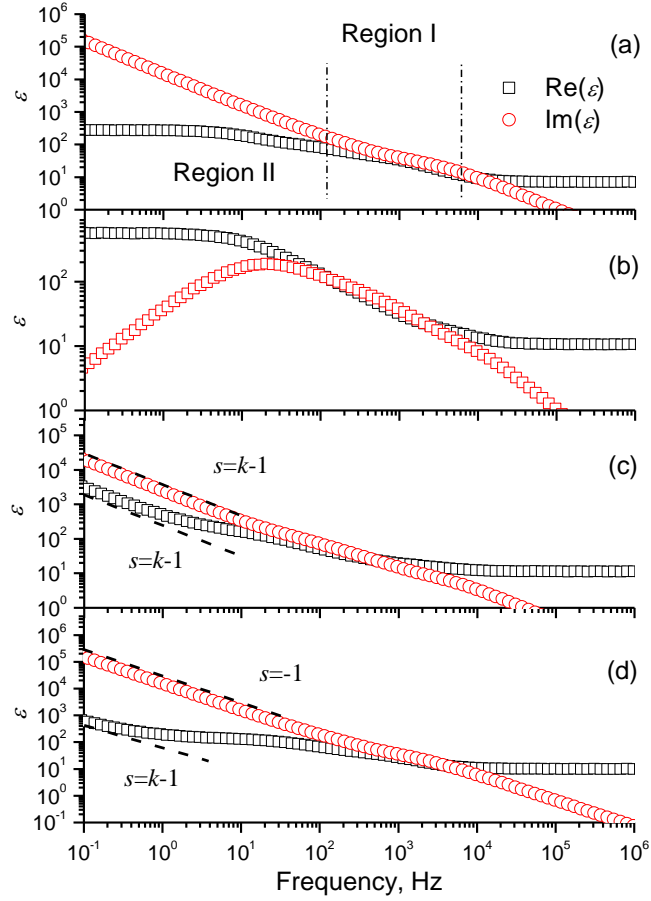

FIG 2: Typical effective permittivity spectra of random RCD networks for different cases: (a) below D percolation and above R percolation, (b) below R and D percolations, (c) below R percolation and above D percolation, (d) above R and D percolations. Here $m=n=5$, $R=1000$ k$\Omega$, $C_0=0.1$ nF, $C_1=50$ nF, $k=0.1$.

Several authors used random RC network simulations for describing the LDF in systems with hopping conducting.[13-15] With such a model, these authors were able to interpret the decrease with frequency of the real and imaginary parts of the effective permittivity of a investigated material with a heterogeneous microstructure at low frequencies (notably in region I in Fig. 2(a)). But region I is also observed in the permittivity spectra below and above the R and D percolations (Fig. 2(b)-(d)). Here R percolation means that it is possible to connect left and right contacts through resistors in series, D percolation means that it is possible to connect left and right contacts through diodes in series or both resistors and diodes. Note that the dielectric dispersion in Fig. 2 (b) is absolutely different in comparison with dielectric dispersion in Fig. 2 (a), (c) and (d). This suggests that another interpretation is needed for a qualitative description of the effective permittivity spectrum of SWCNT-filled polymers for both below and above the percolation threshold. We suggest that the region I is related to dipolar relaxation in composites.[4,21] Additionally, in random RC network simulations a decrease with frequency only of the imaginary part of $\varepsilon$ is observed for percolative systems at low frequencies (Region II). This decrease is associated with finite dc conductivity of the system. We also stress that the introduction of diodes in the system affects significantly the permittivity spectra and allows us to observe LFD in supercolative systems.

Of course, there are many possible configurations of random RCD networks. Polymer composites even with low SWCNT content always possess C percolation paths. The most important situations of random RCD networks that we consider are: (1) no R and D percolation paths, (2) just D percolation path, and (3) both R and D percolation paths present in the system. Note that the case for which there is R percolation path but no D path (Fig. 2a) is unlikely for SWCNT-filled polymers. Let us consider the first situation. The corresponding spectrum of $\varepsilon$ is shown in Fig. 2(b). We observe a typical dielectric response for CNT-filled polymers below percolation threshold which can be described in terms of a quasi-Debye dipolar relaxation.[21,19] The second situation mentioned above is especially important for composites with filler content close to percolation threshold. In this case, LFD is observed (see Fig. 2(c)). The slope $s$ of both real and imaginary parts of $\varepsilon$ is exactly equal to $k-1$ at low frequencies, where $k$ denotes the exponent used for modeling of the diode impedance. The third situation is useful for describing of supercolative composites (Fig. 2(d)). In this case, the slope of the real part of $\varepsilon$ is also equal to $k-1$, but the corresponding slope of the imaginary part of $\varepsilon$ is -1 because the system is conductive. This is in coherence with studies of supercolative SWCNT-filled polymer samples for which such a slope is ranging from -0.8 to -1.[21,10]

This methodology was used to fit experimental data of the permittivity spectra of SWCNT filled composites.[21] The filler particles weight fraction within the composite was denoted as $\varphi$. The composite samples contain $\varphi = 1.5$, 1 and 0.25 wt.% of SWCNT (see Fig. 3(a)). All samples are supercolative, the sample containing 0.25 wt.% of SWCNT is close to the percolation threshold.

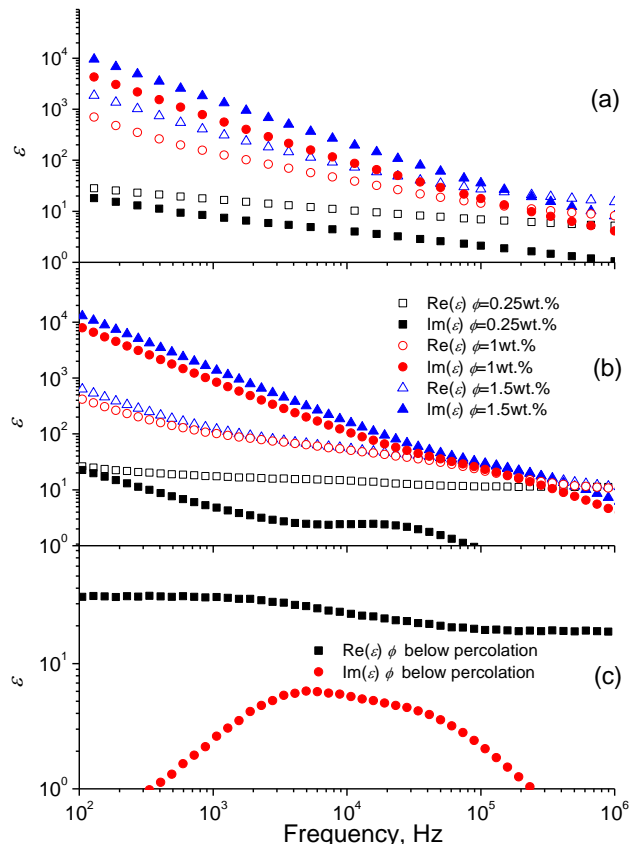

FIG. 3: (a) Experimentally measured effective permittivity spectra of polymer composites containing $\varphi$=1.5, 1 and 0.25 wt.% of SWCNT (reprinted with permission from Ref. [21], copyright 2013 American Institute of Physics); (b) Numerical data obtained from the fitting of the experimental data with the random RCD network model; (c) The predicted response of the SWCNT filled polymer below the percolation threshold using the random RCD network model when there are no D and R percolation paths.

Modeling of these permittivity spectra for samples with 1 and 1.5 wt.% was done by considering a random RCD network ($m=n=5$) in the special situation when it contains R, C and D percolation paths. The fitting parameters are listed in Table I. Our simulations shown in Fig. 3(b) reproduce well the experimental results of Fig. 3(a). For sample containing 0.25 wt.%, the modeling network contains C and D percolation paths but no conductive R paths. This is somewhat intuitive because only one third of the SWCNTs is conductive while the rest is semiconducting.[17] The decrease of SWCNT concentration leads to corresponding increase of resistance R and decrease of the scaling parameter $C_1$ (see Table I). We also used a random RCD network without D and R percolation paths to model the electromagnetic response of composite samples below the percolation threshold (Fig.3(c)). In this case, we observe that decreasing the SWCNT content has for effect to decrease significantly $C_1$. No experimental data were available for the current series of SWCNT-filled samples below the percolation threshold. However, we emphasize that the results obtained are consistent with our previously reported measurements of the permittivity of multiwalled CNT filled polymers.[21]

**Table I:** Best-fit parameters of randomly positioned RCD network obtained with fitting experimental data (from Ref. [21])

| CNT content (wt.%) | $R_0$ (k$\Omega$) | $C_0$ (nF) | $C_1$ ($\mu$F) | k |
|---|---|---|---|---|
| 1.5 | 100 | 0.1 | 1 | 0.1 |
| 1 | 200 | 0.1 | 1 | 0.1 |
| 0.25 | 1000 | 0.1 | $2\times10^{-3}$ | 0.1 |
| below percolation threshold | 1000 | 0.1 | $2\times10^{-5}$ | 0.1 |

To summarize, our results demonstrate that suitably designed random RCD networks provide a means for investigating the quasistatic dielectric response of SWCNT filled polymers. This analytical study allows us to obtain a satisfactory fit of the experimental effective permittivity data. The above-proposed analysis leads to three main conclusions. Firstly, in CNT composites the percolation is complex: at lower CNT concentrations it occurs only in the semiconducting CNT subsystem while at larger CNT content of CNT concentration it occurs also in the metallic CNT subsystem. Secondly, a simple model of a random RCD network is able to describe the two salient features of the dielectric properties of SWCNT filled composites, i.e. the LDF and dipolar relaxation. Thirdly, the LFD in supercolative composite samples is primarily caused by the fraction of semiconducting SWCNTs. Overall, there are hints here of a general phenomenon: the emergence of collective behavior within complex networks which has been observed in materials as diverse as carbonaceous materials[4] and amorphous materials.[13-15,19,20]

**Acknowledgements**

The work was financially supported by the RFBR (Grant No. 12-02-90011), EU Projects PIRSES-2012-318617 FAEMCAR, and PIRSES-GA-2013-610875 NAmiceMC, and BRFBR project FR12-030 and the European Social Fund under the Global Grant measure. Lab-STICC is UMR CNRS 6285.

[a]dzmitrybychanok@yandex.by

**References**

[1] S. El Bouazzaoui, A. Droussi, M. E. Achour, and C. Brosseau, J. Appl. Phys. **106**, 104107 (2009).

[2] L. Ren, C. L. Pint, L. G. Booshehri, W. D. Rice, X. Wang, D. J. Hilton, K. Takeya, I. Kawayama, M. Tonouchi, R. H. Hauge, and J. Kono, Nano Lett. **9**, 2610 (2009).

[3] L. Licea-Jimenez, A. D. Grishina, L.Y. Pereshivko, T. V. Krivenko, V. V. Savelyev, R. W. Rychwalski, and A. V. Vannikov, Carbon **44**, 113 (2006).

[4] F. Qin and C. Brosseau, J. Appl. Phys. **111**, 061301 (2012).

[5] A. V. Okotrub, V. V. Kubarev, M. A. Kanygin, O. V. Sedelnikova, and L. G. Bulusheva, Phys. Status Solidi B, 248, 2568 (2011).

[6] D. Bychanok, M. Kanygin, A. Okotrub, M. Shuba, A. Paddubskaya, A. Pliushch, P. Kuzhir, and S. Maksimenko, JETP Lett. **93**, 607 (2011).

[7] J. Wu and L. Kong, Appl. Phys. Lett. **84**, 4956 (2004).

[8] D. S. Bychanok, M. V. Shuba, P. P. Kuzhir, S. A. Maksimenko, V. V. Kubarev, M. A. Kanygin, O. V. Sedelnikova, L. G. Bulusheva, and A. V. Okotrub, J. Appl. Phys. **114**, 114304 (2013).

[9] C. Brosseau, Surf. Coating Tech. **206**, 753 (2011).

[10] M. E. Achour, C. Brosseau, and F. Carmona, J. Appl. Phys. **103**, 094103 (2008).

[11] C. Brosseau, F. Boulic, P. Queffelec, C. Bourbigot, Y. Le Mest, J. Loaec, and A. Beroual, J. Appl. Phys. **81**, 882 (1997).

[12] C. Brosseau, P. Molinie, F. Boulic, and F. Carmona, J. Appl. Phys. **89**, 8297 (2001).

[13] D. P. Almond and B. Vainas, J. Physics: Cond. Matter **11**, 9081 (1999).

[14] D. P. Almond, C. R. Bowen, and D. A. S. Rees, J. Phys. D **39**, 1295 (2006).

[15] C. R. Bowen and D. P. Almond, Mater. Sci. Technol. **22,** 719 (2006).

[16] P. Smilauer, Thin Solid Films, **203**, 1 (1991).

[17] R. Saito, G. Dresselhaus, and M. S. Dresselhaus, *Physical Properties of Carbon Nanotubes* (Imperial College Press, London, 1998).

[18] W. Wang, S. Haruehanroengra, L. Shang, M. Liu, Micro & Nano Lett., **2**, 35-39 (2007).

[19] A. K. Jonscher, *Dielectric Relaxation in Solids* (Chelsea Dielectric Press, London, 1987).

[20] A. K Jonscher, J. Phys. D **32**, R57 (1999).

[21] D. Bychanok, P. Kuzhir, S. Maksimenko, S. Bellucci, and C. Brosseau, J. Appl. Phys. **113**, 124103 (2013).

[22] P. Kuzhir, A. Paddubskaya, D. Bychanok, A. Nemilentsau, M. Shuba, A. Plusch, S. Maksimenko, S. Bellucci, L. Coderoni, F. Micciulla, I. Sacco, G. Rinaldi, J. Macutkevic, D. Seliuta, G. Valusis, J. Banys, Thin Solid Films **519**, 4114 (2011).

[23] M. Donahue, B. Lubbers, M. Kittler, P. Mai, and A. Schober, Appl. Phys. Lett. **102**, 141607 (2013).